\documentclass[twocolumn,aps,prl,showpacs,superscriptaddress]{revtex4}
\usepackage{psfig}

\newcommand {\snn}	{\sqrt{s_{_{\rm NN}}}}
\newcommand {\Nch}	{N_{\rm ch}}
\newcommand {\dNch}	{dN_{\rm ch}/d\eta}
\newcommand {\ntrig}	{N_{\rm trig}}
\newcommand {\nassoc}	{\mathcal{N}_{\rm ch}}
\newcommand {\spt}	{\mathcal{P}_{\perp}}
\newcommand {\vpt}	{\vec{\mathcal{P}}_{\perp}}
\newcommand {\dca}	{dca}
\newcommand {\pt}	{p_{\perp}}
\newcommand {\pttrig}	{p_{\perp}^{\rm trig}}
\newcommand {\mpt}	{\langle p_{\perp} \rangle}
\newcommand {\mpttrig}	{\langle p_{\perp}^{\rm trig} \rangle}
\newcommand {\deta}	{\Delta\eta}
\newcommand {\dphi}	{\Delta\phi}
\newcommand {\vflow}	{v_2}
\newcommand {\vMRP}	{$v_2$\{{\sc mrp}\}}
\newcommand {\vCUM}	{$v_2$\{4\}}
\newcommand {\Iaa}	{I_{AA}}

\begin{document}


\title{Distributions of Charged Hadrons Associated with High Transverse Momentum Particles in $pp$ and Au+Au Collisions at $\snn$=200~GeV}

\affiliation{Argonne National Laboratory, Argonne, Illinois 60439}
\affiliation{Brookhaven National Laboratory, Upton, New York 11973}
\affiliation{University of Birmingham, Birmingham, United Kingdom}
\affiliation{University of California, Berkeley, California 94720}
\affiliation{University of California, Davis, California 95616}
\affiliation{University of California, Los Angeles, California 90095}
\affiliation{California Institute of Technology, Pasadena, California 91125}
\affiliation{Carnegie Mellon University, Pittsburgh, Pennsylvania 15213}
\affiliation{Creighton University, Omaha, Nebraska 68178}
\affiliation{Nuclear Physics Institute AS CR, \v{R}e\v{z}/Prague, Czech Republic}
\affiliation{Laboratory for High Energy (JINR), Dubna, Russia}
\affiliation{Particle Physics Laboratory (JINR), Dubna, Russia}
\affiliation{University of Frankfurt, Frankfurt, Germany}
\affiliation{Indian Institute of Technology, Mumbai, India}
\affiliation{Indiana University, Bloomington, Indiana 47408}
\affiliation{Institute  of Physics, Bhubaneswar 751005, India}
\affiliation{Institut de Recherches Subatomiques, Strasbourg, France}
\affiliation{University of Jammu, Jammu 180001, India}
\affiliation{Kent State University, Kent, Ohio 44242}
\affiliation{Lawrence Berkeley National Laboratory, Berkeley, California 94720}\affiliation{Max-Planck-Institut f\"ur Physik, Munich, Germany}
\affiliation{Michigan State University, East Lansing, Michigan 48824}
\affiliation{Moscow Engineering Physics Institute, Moscow Russia}
\affiliation{City College of New York, New York City, New York 10031}
\affiliation{NIKHEF, Amsterdam, The Netherlands}
\affiliation{Ohio State University, Columbus, Ohio 43210}
\affiliation{Panjab University, Chandigarh 160014, India}
\affiliation{Pennsylvania State University, University Park, Pennsylvania 16802}
\affiliation{Institute of High Energy Physics, Protvino, Russia}
\affiliation{Purdue University, West Lafayette, Indiana 47907}
\affiliation{University of Rajasthan, Jaipur 302004, India}
\affiliation{Rice University, Houston, Texas 77251}
\affiliation{Universidade de Sao Paulo, Sao Paulo, Brazil}
\affiliation{University of Science \& Technology of China, Anhui 230027, China}
\affiliation{Shanghai Institute of Nuclear Research, Shanghai 201800, P.R. China}
\affiliation{SUBATECH, Nantes, France}
\affiliation{Texas A\&M University, College Station, Texas 77843}
\affiliation{University of Texas, Austin, Texas 78712}
\affiliation{Valparaiso University, Valparaiso, Indiana 46383}
\affiliation{Variable Energy Cyclotron Centre, Kolkata 700064, India}
\affiliation{Warsaw University of Technology, Warsaw, Poland}
\affiliation{University of Washington, Seattle, Washington 98195}
\affiliation{Wayne State University, Detroit, Michigan 48201}
\affiliation{Institute of Particle Physics, CCNU (HZNU), Wuhan, 430079 China}
\affiliation{Yale University, New Haven, Connecticut 06520}
\affiliation{University of Zagreb, Zagreb, HR-10002, Croatia}
\author{J.~Adams}\affiliation{University of Birmingham, Birmingham, United Kingdom}
\author{C.~Adler}\affiliation{University of Frankfurt, Frankfurt, Germany}
\author{M.M.~Aggarwal}\affiliation{Panjab University, Chandigarh 160014, India}
\author{Z.~Ahammed}\affiliation{Variable Energy Cyclotron Centre, Kolkata 700064, India}
\author{J.~Amonett}\affiliation{Kent State University, Kent, Ohio 44242}
\author{B.D.~Anderson}\affiliation{Kent State University, Kent, Ohio 44242}
\author{D.~Arkhipkin}\affiliation{Particle Physics Laboratory (JINR), Dubna, Russia}
\author{G.S.~Averichev}\affiliation{Laboratory for High Energy (JINR), Dubna, Russia}
\author{S.K.~Badyal}\affiliation{University of Jammu, Jammu 180001, India}
\author{J.~Balewski}\affiliation{Indiana University, Bloomington, Indiana 47408}
% ...

\collaboration{STAR Collaboration}\homepage{www.star.bnl.gov}\noaffiliation

\begin{abstract}
Charged hadrons in 0.15$<$$\pt$$<$4~GeV/$c$ associated with particles of $\pttrig$$>$4~GeV/$c$ are reconstructed in $pp$ and Au+Au collisions at $\snn$=200 GeV. The associated multiplicity and $\pt$ magnitude sum are found to increase from $pp$ to central Au+Au collisions. The associated $\pt$ distributions, while similar in shape on the near side, are significantly softened on the away side in central Au+Au relative to $pp$ and not much harder than that of inclusive hadrons. The results, consistent with jet quenching, suggest that the awayside fragments approach equilibration with the medium traversed.
\end{abstract}
\maketitle

Quantum Chromodynamics (QCD) predicts a phase transition between hadronic matter and quark-gluon plasma at a critical energy density of $\sim$1~GeV/fm$^3$~\cite{lattice}. Such a phase transition is being actively pursued at the Relativistic Heavy-Ion Collider (RHIC). High transverse momentum ($\pt$) particles, emerging from hard scatterings, lose energy while traversing and interacting with the medium being developed in heavy-ion collisions. Energy loss results in jet quenching~\cite{jetQuenching} -- suppressions of hadron yield and back-to-back angular correlation at high $\pt$. Such suppressions were observed in central Au+Au collisions at RHIC~\cite{suppression,back2back} and attributed to final state interactions when no suppression was seen in d+Au~\cite{dAu}. Perturbative QCD model calculations invoking parton energy loss require 30 times the normal nuclear gluon density in order to account for the central Au+Au results~\cite{density}.

The depleted energy at high $\pt$ must be redistributed to low $\pt$ particles~\cite{theory,theory2}. Reconstruction of these particles will constrain models describing production mechanisms of high $\pt$ particles, and may shed light on the underlying energy loss mechanism(s) and the degree of equilibration of jet products with the medium. 

This Letter presents results from statistical reconstruction, via two-particle angular correlations, of charged hadrons in 0.15$<$$\pt$$<$4~GeV/$c$ associated with a high $\pt$ ``trigger" particle in $pp$ and Au+Au collisions at $\snn$=200~GeV. Two $\pt$ windows for trigger particles, 4$<$$\pttrig$$<$6~GeV/$c$ and 6$<$$\pttrig$$<$10~GeV/$c$, are presented. The latter range is expected~\cite{coalescence,Hwa} to provide a purer, though much lower statistics, sample of hard scattering products. Results are reported as a function of centrality for Au+Au collisions and the associated hadron $\pt$.

{\em Analysis.--} 
The STAR experiment~\cite{StarNIM} is well suited for this measurement due to significant pseudo-rapidity ($\eta$) and complete azimuthal ($\phi$) coverage. The STAR Time Projection Chamber (TPC) resides in a magnetic field of 0.5 T along its cylindrical axis (= the beam direction). Events with reconstructed primary vertex within $\pm 25$ cm longitudinally of the TPC center are used. The Au+Au events are divided into 7 centrality classes as in~\cite{back2back}. 

High $\pt$ trigger particles are selected with $|\eta_{\rm trig}|$$<$0.7 and $\dca$ (distance of closest approach to the primary vertex) $<$1 cm. Other particles in the event with $|\eta|$$<$1.0 and $\dca$$<$2 cm are paired with each trigger particle to form $\deta$=$\eta$$-$$\eta_{\rm trig}$ and $\dphi$=$\phi$$-$$\phi_{\rm trig}$ distributions. The primary vertex is included in the momentum fit of the associated particles, but not for trigger particles to minimize weak decay background. 

Combinatorial coincidences are removed by subtracting mixed-event background of the same centrality bin, so that detector non-uniformities should affect signal and background distributions in the same way. The effect of elliptic flow ($\vflow$) is included by multiplying the Au+Au mixed-event background by $1+2\vflow(\pttrig)\vflow(\pt)\cos(2\dphi)$~\cite{Kirill}. The mixed events may not precisely match the underlying background in events with a trigger particle, {\em e.g.}, due to different centrality distributions within each analyzed bin. Hence, an additional $\pt$-independent factor (1.46 for $pp$ and 0.995-1.000 for Au+Au) has been applied to the background before subtraction, in order to normalize it to the measured $\dphi$ distribution within 0.8$<$$|\dphi|$$<$1.2 for 0.15$<$$\pt$$<$4 GeV/$c$.

\begin{figure}[hbt]
\vspace*{-4mm}
\centerline{\hspace*{8mm}\psfig{file=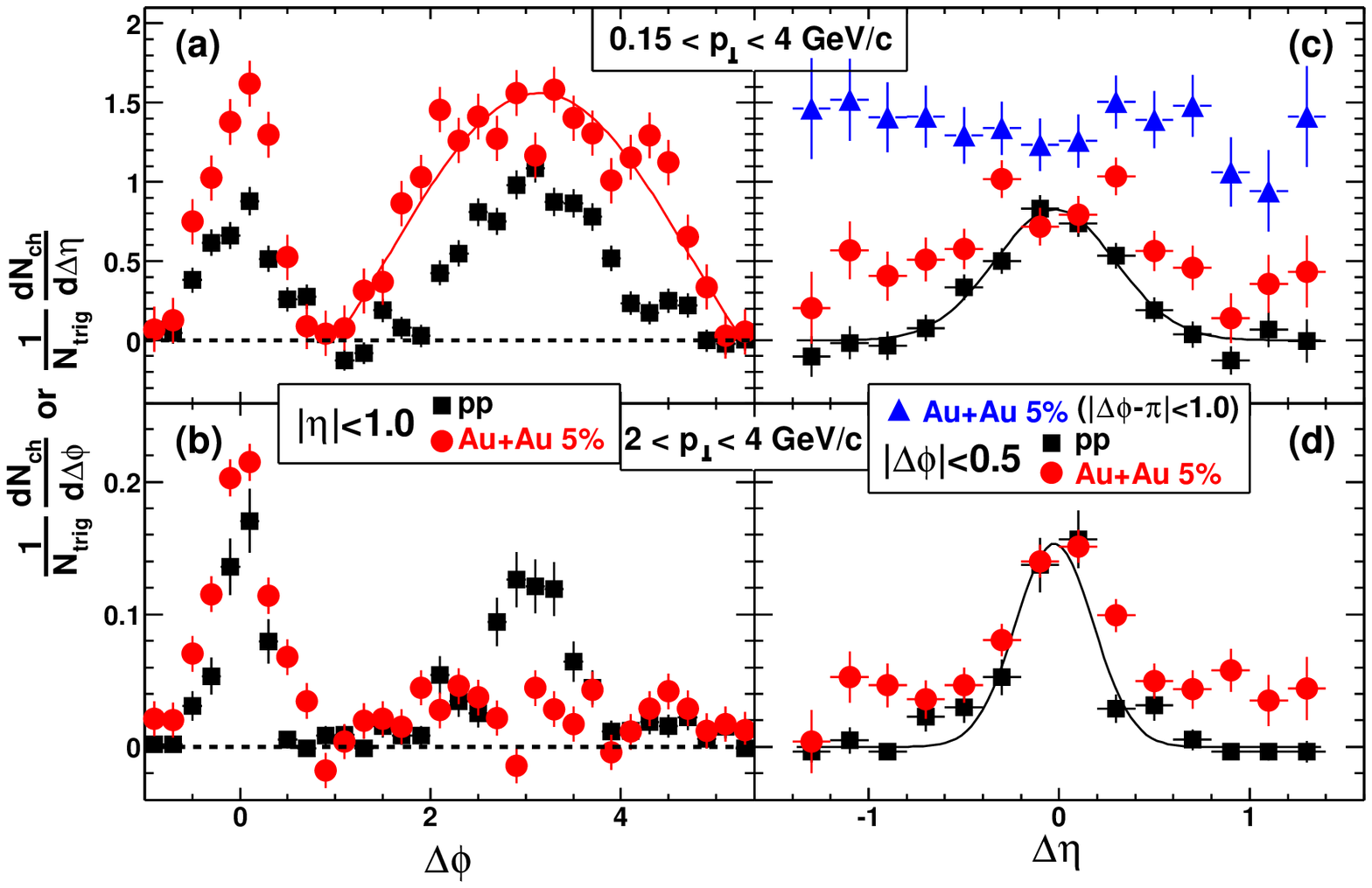,width=0.53\textwidth}}
\vspace{-4mm}
\caption{Background subtracted (a),(b) $\dphi$ and (c),(d) $\deta$ distributions for $pp$ and 5-0\% central Au+Au for 4$<$$\pttrig$$<$6~GeV/$c$ and two associated $\pt$ ranges. The subtracted background level for $\pt$=0.15-4~GeV/$c$ (2-4~GeV/$c$) is $\frac{1}{\ntrig}\frac{d\Nch}{d\Delta\phi}\approx$1.4 (0.007) in $pp$ and $\approx$211 (2.1) in 5-0\% Au+Au. The curve in (a) shows the shape of an $[A-B\cos(\dphi)]$ function. The curves in (c),(d) are Gaussian fits to the $pp$ data.}
\label{fig1}
\end{figure}

Figure~\ref{fig1} compares the background-subtracted $\dphi$ and $\deta$ distributions for $pp$ vs central Au+Au collisions, including [\ref{fig1}(a) and \ref{fig1}(c)] or excluding [\ref{fig1}(b) and \ref{fig1}(d)] the softest associated particles. The distributions are corrected for single-particle (and, in the case of $\deta$, for two-particle) acceptance and efficiency, and are normalized per detected trigger particle. The $\dphi$ distributions in \ref{fig1}(b) support the qualitative conclusions of~\cite{back2back}, exhibiting near- ($\dphi$$\approx$0) and awayside ($\dphi$$\approx$$\pi$) jet peaks, with the latter strongly suppressed by jet quenching in central Au+Au. Comparison of \ref{fig1}(a) and \ref{fig1}(b) shows that more soft associated hadrons are found per trigger particle in central Au+Au than in $pp$, on both the near and away sides. Inclusion of the soft particles broadens the $\dphi$ peaks, especially on the away side. Indeed, the awayside strength for central Au+Au in \ref{fig1}(a) is no longer even "jet"-like, but is rather consistent in shape with the $[A-B\cos(\dphi)]$ dependence expected~\cite{borghini} for purely statistical momentum balance of the nearside jet.

For associated hadrons within the nearside $\dphi$ region, the $\deta$ distributions shown in Fig.~\ref{fig1}(c) and \ref{fig1}(d) exhibit jet-like peaks that are broader for central Au+Au than for $pp$, and grow broader still in both cases when the soft associated hadrons are included. The awayside hadrons have an essentially flat distribution in $\deta$ over the measured range for both $pp$ and Au+Au - the latter are shown in \ref{fig1}(c) - as expected when a broad range of parton momenta contribute to jet production. This flat $\deta$ distribution, combined with the limited TPC coverage ($|\deta|$$<$1.4), implies that we cannot hope to recover the full awayside momentum needed to balance the nearside jets.

To accommodate the features in Fig.~\ref{fig1}, we define nearside ($|\dphi|$$<$1.0, $|\deta|$$<$1.4) and awayside ($|\dphi|$$>$1.0, $|\eta|$$<$1.0) regions for the remaining analysis. We integrate the correlation peaks as measures of associated charged hadron multiplicities ($\nassoc$). We obtain $\pt$ magnitude sum ($\spt$=$\sum{\pt}$), which approximates associated energy, and vector {sum} ($\vpt$=$\sum{\pt\cos\dphi}$) from the $\pt$-weighted $\dphi$ and $\deta$ distributions multiplied by $1.58\pm 0.08$~\cite{note_neutral} to account for the undetected neutrals. The $\mpttrig$ is then added in $\spt$ and $\vpt$ for the near side.

{\em Systematic errors.--} 
Table~\ref{sys} lists the major sources of systematic uncertainties in $\nassoc$. (1) The acceptance and efficiency correction has a 10\% uncertainty. (2) In constructing background, we use the average of the $\vflow$ results from the modified reaction plane (\vMRP)~\cite{v2PRC} and 4-particle (\vCUM)~\cite{v2_4part} methods and assign the difference as uncertainty. For the 80-60\% and 5-0\% centralities where \vCUM\ are unavailable, we estimate \vCUM$\approx$\vMRP/2. Relatively small uncertainties arise on the away side because the $\dphi$ integration range is much broader than $\pi/2$ and the background normalization is correlated with the $\vflow$ correction used. (3) Uncertainties in background normalization for 0.15$<$$\pt$$<$4 GeV/$c$ are estimated by varying the $\dphi$ region for normalization. (4) An additional (single-sided) uncertainty due to possible $\pt$-dependent differences between the mixed-event and true background is estimated by comparing to results using $\pt$-dependent background normalization. The systematic errors from the above sources are added in quadrature, separately for the positive and negative uncertainties.

\begin{table}[hbt]
\caption{Major sources of systematic uncertainties (in percent) in $\nassoc$ for 4$<$$\pttrig$$<$6~GeV/$c$.}
\label{sys}
\begin{ruledtabular}
\begin{tabular}{l|cc|cc|cc|cc}
 & \multicolumn{2}{c|}{$pp$} & \multicolumn{2}{c|}{80-60\%} & \multicolumn{2}{c|}{30-20\%} & \multicolumn{2}{c}{5-0\%} \\
\multicolumn{1}{c|}{source} & near & away  & near & away  & near & away & near & away \\ \hline
(1) effic. & \multicolumn{2}{c|}{$\pm 10$} & \multicolumn{2}{c|}{$\pm 10$} & \multicolumn{2}{c|}{$\pm 10$} & \multicolumn{2}{c}{$\pm 10$} \\
(2) flow & \multicolumn{2}{c|}{--} & $^{+34}_{-40}$ & $\pm 4$ & $^{+21}_{-22}$ & $\pm 5$ & $^{+19}_{-27}$ & $^{+4}_{-5}$ \\ 
(3) bkgd. & $^{+22}_{-13}$ & $^{+22}_{-14}$ & $^{+62}_{-6}$ & $^{+36}_{-4}$ & $^{+27}_{-12}$ & $^{+32}_{-14}$ & $^{+11}_{-14}$ & $^{+10}_{-13}$ \\ \hline
\multicolumn{1}{c}{} & \multicolumn{2}{r|}{\hspace{-5mm}$\pt$(GeV/$c$)} & \multicolumn{2}{c|}{$pp$} & \multicolumn{2}{c|}{80-40\%} & \multicolumn{2}{c}{5-0\%} \\ \cline{2-9}
\multicolumn{1}{l}{(4) $\pt$--} & \multicolumn{2}{c|}{0.5-1.0} & $ +1$ & $ -7$ & $-39$ & $ -6$ & $ -5$ & $ +1$ \\
\multicolumn{1}{l}{\hspace*{4mm} dep.} & \multicolumn{2}{c|}{1.5-2.0} & $-25$ & $-29$ & $-28$ & $-25$ & $ +7$ & $ +7$ \\
\multicolumn{1}{l}{} & \multicolumn{2}{c|}{2.5-3.0} & $ -1$ & $-16$ & $ -9$ & $-16$ & $ -6$ & $-15$ \\
\end{tabular}
\end{ruledtabular}
\end{table}

\begin{figure}[hbt]
\vspace*{-4mm}
\centerline{\psfig{file=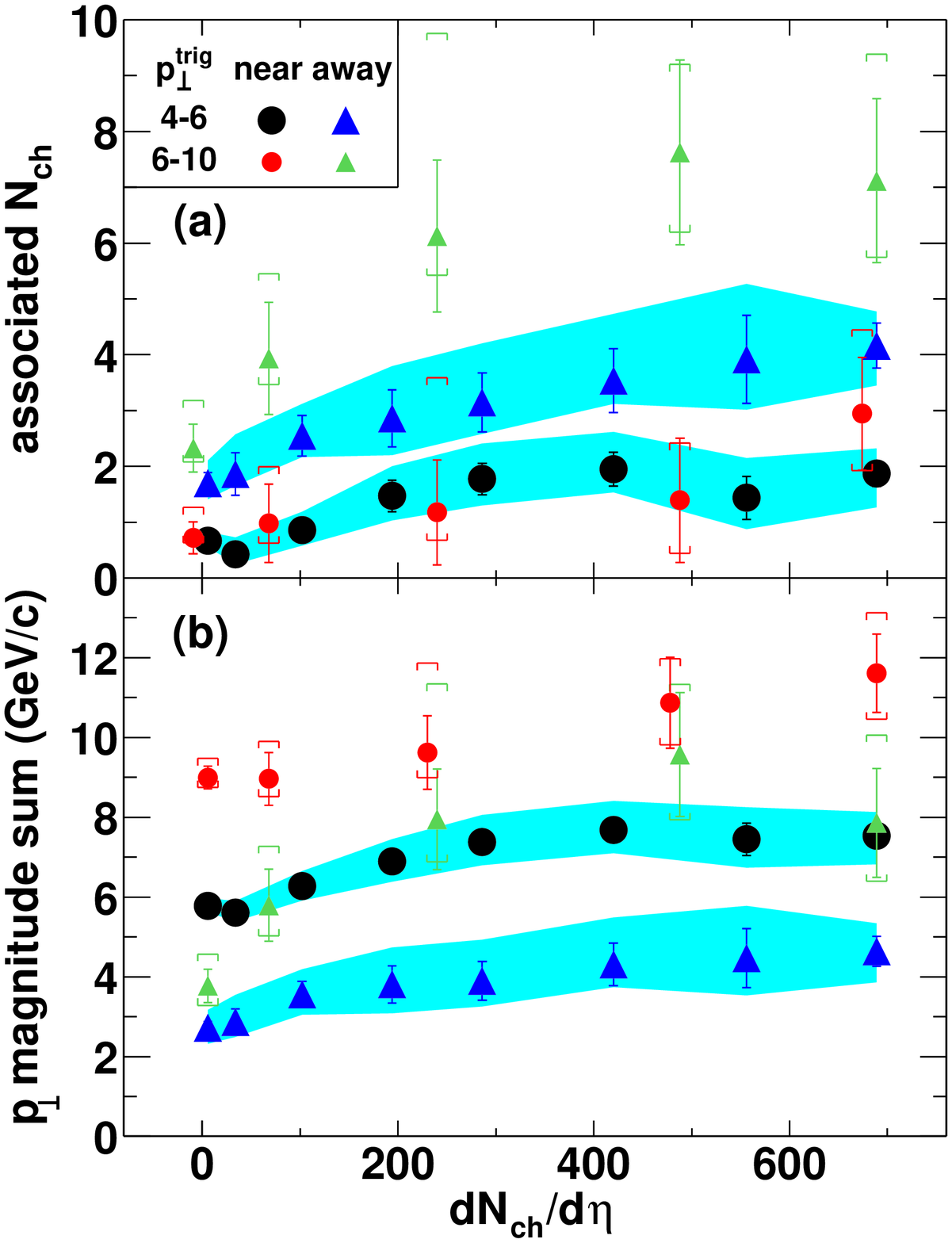,width=0.4\textwidth}}
\vspace*{-4mm}
\caption{(a) $\nassoc$ and (b) $\spt$ for $\pttrig$=4-6 (6-10)~GeV/$c$ with systematic errors in bands (caps). Systematic errors are strongly correlated between near and away side and among the centralities. The leftmost set of data are for $pp$. Some of the open points are slightly displaced in $\dNch$ for clarity.}
\label{fig2}
\end{figure}

{\em Results.--} 
Figure~\ref{fig2} shows $\nassoc$ and $\spt$ in $pp$ and as a function of centrality (the charged hadron $\dNch$) in Au+Au collisions for the two $\pttrig$ windows of 4-6~GeV/$c$ and 6-10~GeV/$c$. For $pp$ and all centralities of Au+Au, $\mpttrig$$\approx$4.55~GeV/$c$ and 7.0~GeV/$c$ for the two $\pttrig$ windows, respectively. With the same $\mpttrig$ trigger particle, $\nassoc$ and $\spt$ increase from $pp$ to central Au+Au for both the near and away sides, and for both $\pttrig$ selections.

Our results include nearly all associated hadrons on the near side but, as noted above, only the fraction within our acceptance on the away side. We find the away to near side $|\vpt|$ ratio$\approx$40\%, independent of system or centrality.

\begin{figure}[hbt]
\centerline{\psfig{file=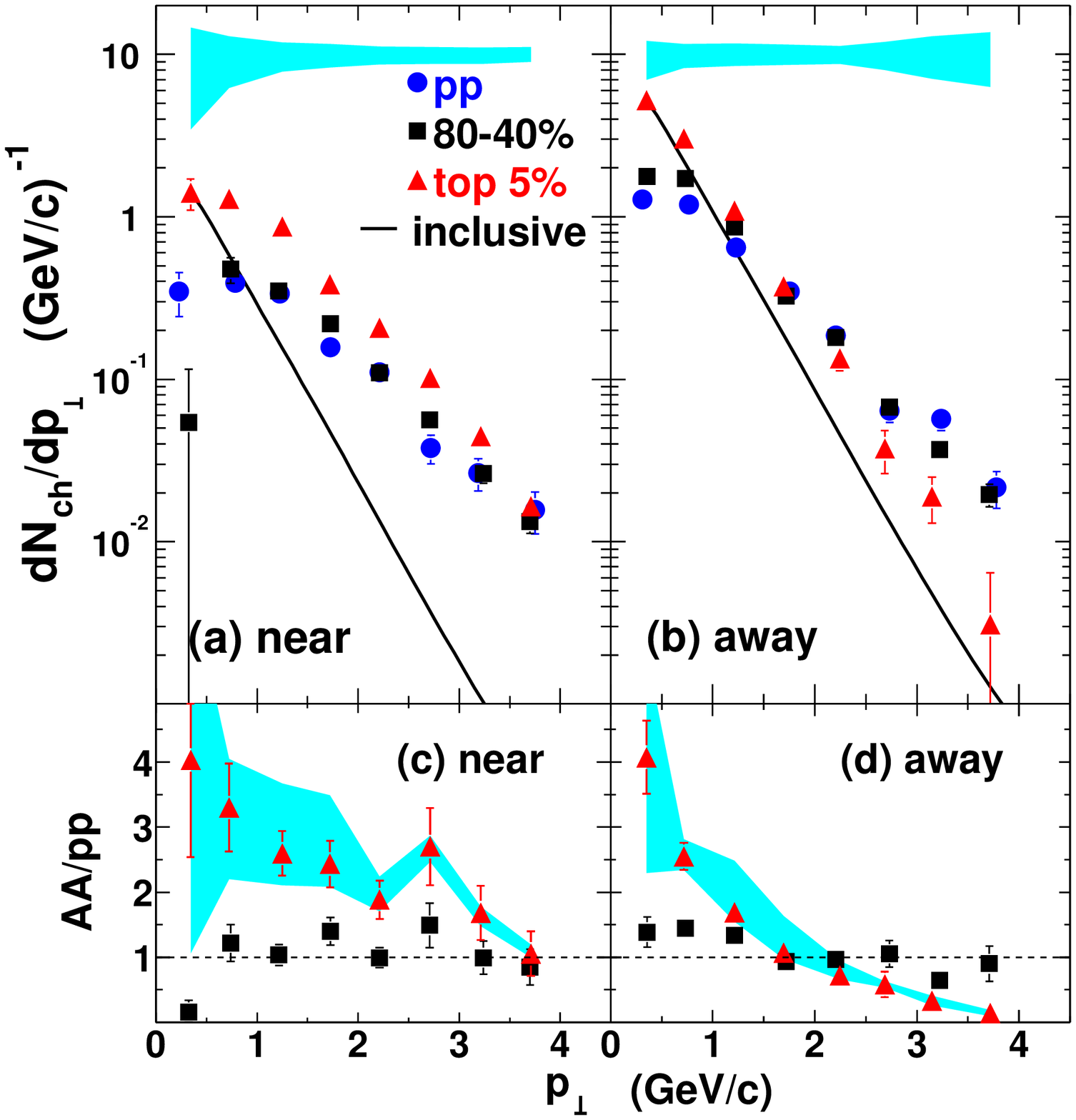,width=0.48\textwidth}}
\vspace*{-6mm}
\caption{Associated charged hadron $\pt$ distributions (a),(b) and AA/$pp$ ratios (c),(d) for 4$<$$\pttrig$$<$6 GeV/$c$ on near and away side. Errors shown are statistical. The bands show the systematic uncertainties for the 5-0\% central data. The lines show the inclusive spectral shape for central collisions.}
\label{fig3}
\end{figure}

Figure~\ref{fig3}(a) and~\ref{fig3}(b) shows the $\pt$ distributions of associated charged hadrons for 4$<$$\pttrig$$<$6~GeV/$c$ in $pp$, peripheral 80-40\% and central 5-0\% Au+Au collisions. The Au+Au to $pp$ spectra ratios (AA/$pp$) are depicted in Fig.~\ref{fig3}(c) and~\ref{fig3}(d). In the systematic uncertainties for AA/$pp$, sources (1), (3), and (4) in Table~\ref{sys} tend to cancel. Results for peripheral Au+Au generally agree with $pp$ (AA/$pp$$\approx$1), while those for central Au+Au differ. On the near side, the central Au+Au collisions show a larger multiplicity of associated hadrons, but with $\mpt$=1.02$\pm$0.05(stat)$^{+0.17}_{-0.08}$(syst)~GeV/$c$ essentially unchanged within uncertainties from its $pp$ value (1.15$\pm$0.06$^{+0.14}_{-0.17}$~GeV/$c$). On the away side, the spectrum is significantly softened in central Au+Au collisions; associated particles are depleted at high $\pt$, as first noted in~\cite{back2back}, and are significantly enhanced at low $\pt$.

AA/$pp$ cannot be readily compared to the analogous $\Iaa$ defined in~\cite{back2back}, due to differences in integration ranges and methodology: {\em e.g.}, the $\Iaa$ prescription in~\cite{back2back} omits two-particle acceptance corrections, and thereby suppresses long-range $\deta$ correlation signals which may contribute to AA/$pp$ after mixed-event and elliptic flow subtractions. To permit quantitative comparison, we also extract $\Iaa$ using the same procedures and momentum range (2$<$$\pt$$<$4~GeV/$c$) as in~\cite{back2back}. The extracted values for 80-60\% Au+Au are 0.99$\pm$0.11(stat)$^{+0.06}_{-0.08}$(syst) and 0.85$\pm$0.09$^{+0.05}_{-0.07}$ for near and away side, respectively; those for 5-0\% Au+Au are 1.55$\pm$0.14$^{+0.13}_{-0.17}$ and 0.28$\pm$0.06$^{+0.10}_{-0.14}$. They differ numerically from \cite{back2back} primarily due to our use of reduced $\vflow$ values and a more stringent primary vertex cut for $pp$. The systematic errors quoted for $\Iaa$ are from $\vflow$ and background uncertainties, the latter estimated by fitting to observed $\dphi$ distributions over several ranges beyond the default 0.75$<$$\dphi$$<$2.24 used in~\cite{back2back}.

\begin{figure}[hbt]
\vspace*{-3mm}
\centerline{\psfig{file=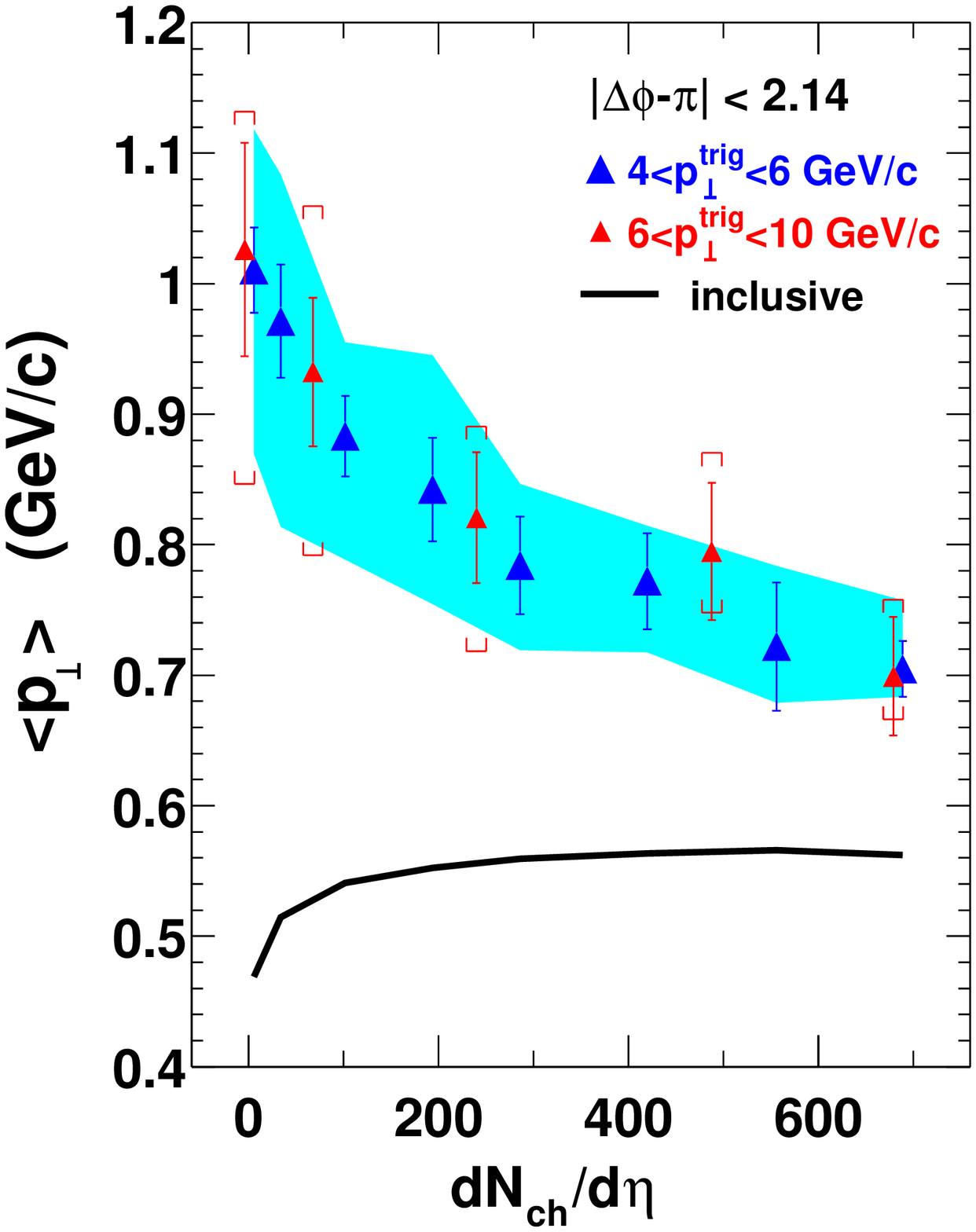,width=0.4\textwidth}}
\vspace*{-14mm}
\caption{Awayside associated hadron $\mpt$ for $\pttrig$=4-6 (6-10)~GeV/$c$ with systematic errors in band (caps).}
\label{fig4}
\end{figure}

Figure~\ref{fig4} shows the centrality dependence of $\mpt$ of the awayside associated hadrons. For both $\pttrig$ selections, $\mpt$ drops rapidly with increasing centrality, while that of {\em inclusive} hadrons (i.e. without trigger particle selection, in curve) rises. The trend toward convergence of the $\mpt$ for these two samples may indicate a progressive equilibration of the awayside associated hadrons with the bulk medium from peripheral to central collisions. 

{\em Discussion.--} 
High $\pt$ hadrons arise predominantly from jets in $pp$ and peripheral Au+Au collisions~\cite{jetQuenching}, but softer production mechanisms~\cite{coalescence,Hwa} may be of comparable importance in central Au+Au for 4$<$$\pttrig$$<$6~GeV/$c$. Such softer contributions are expected to be negligible in the 6$<$$\pttrig$$<$10~GeV/$c$ region. The consistency between the two $\pttrig$ windows thus suggests that the results reflect features of hard scattering in Au+Au collisions.

In the context of hard parton scattering and subsequent energy loss, high $\pttrig$ particles select preferentially di-jets produced near the medium surface~\cite{back2back}. The nearside jet traverses and interacts with a minimal amount of matter. No broadening relative to $pp$ is observed for the nearside $\dphi$ correlation. The observed broadening in $\deta$ is possibly due to transverse and/or longitudinal flow of the medium~\cite{voloshin}. More hadrons and energy accompany the same $\mpttrig$ trigger particle in central Au+Au than in $pp$. This could be the net effect of modest parton energy loss softening the resulting jet fragmentation function~\cite{theory2}, plus energy pickup from the medium, part of which becomes correlated with the trigger through processes such as recombination~\cite{Hwa}, scattering, or flow~\cite{voloshin}.

The awayside jet traverses a large amount of matter. Significant energy loss occurs, depleting high $\pt$ and enhancing low $\pt$ fragments. Energy transferred from high to low $\pt$ results in an increase in the total associated hadron multiplicity. Given the limited TPC acceptance for away jets, our results indicate a large difference between $pp$ and central Au+Au collisions; a significant amount of associated energy may come from the medium in central collisions. The final remnants in central Au+Au no longer exhibit jet-like angular correlations. The interactions seem to drive particles from the two sources, jet fragmentation and the bulk medium, toward equilibration. This may in turn imply a high degree of thermalization within the medium itself.

{\em Conclusions.--} 
We have reported results on statistical reconstruction, via two-particle angular correlations, of charged hadrons in 0.15$<$$\pt$$<$4~GeV/$c$ associated with particles of $\pttrig$$>$4~GeV/$c$ in $pp$ and Au+Au collisions at RHIC. For a given trigger momentum $\mpttrig$, associated hadron multiplicity and $\pt$ magnitude sum increase from $pp$ to central Au+Au collisions. The transverse momentum distributions of associated hadrons, while similar in shape on the near side, are significantly softened on the away side in central Au+Au relative to $pp$. The $\mpt$ of the awayside associated hadrons decreases with centrality, and becomes not much larger than that of inclusive hadrons, indicating a progressive equilibration between the awayside hadrons and the medium. The results are qualitatively the same for 4$<$$\pttrig$$<$6~GeV/$c$ and 6$<$$\pttrig$$<$10~GeV/$c$, and are qualitatively consistent with modification of jets in heavy-ion collisions at RHIC.

We thank the RHIC Operations Group and RCF at BNL, and the NERSC Center at LBNL for their support. This work was supported in part by the HENP Divisions of the Office of Science of the U.S. DOE; the U.S. NSF; the BMBF of Germany; IN2P3, RA, RPL, and EMN of France; EPSRC of the United Kingdom; FAPESP of Brazil; the Russian Ministry of Science and Technology; the Ministry of Education and the NNSFC of China; Grant Agency of the Czech Republic, FOM and UU of the Netherlands, DAE, DST, and CSIR of the Government of India; the Swiss NSF.


\end{document}